\documentclass[conference]{IEEEtran}
\IEEEoverridecommandlockouts
\usepackage{cite}
\usepackage{amsmath,amssymb,amsfonts}
\usepackage{algorithmic}
\usepackage{graphicx}
\usepackage{textcomp}
\usepackage{xcolor}
\usepackage[colorlinks=true,linkcolor=blue,citecolor=blue,urlcolor=blue]{hyperref}
\usepackage[nameinlink,capitalise,noabbrev]{cleveref}

\usepackage{enumitem}
\begin{document}

\title{When the Miner is the Generator: Rethinking the Separation Axiom in LLM-Driven Data Mining}

\author{\IEEEauthorblockN{Xi Chen}
\IEEEauthorblockA{\textit{CSSE, Shenzhen University} \\
Shenzhen, China \\
2024150062@mails.szu.edu.cn}
\and
\IEEEauthorblockN{Mengke Li*}
\IEEEauthorblockA{\textit{CSSE, Shenzhen University} \\
Shenzhen, China \\
mengkeli@szu.edu.cn}
\thanks{* Corresponding author. 

This work was supported in part by NSFC (62676254, 62306181), Guangdong Basic and Applied Basic Research Foundation (2024A1515010163), and National Key Lab of Radar Signal Processing (JKW202403)}
}

\maketitle

\begin{abstract}
The theoretical foundation of data mining implicitly relies on an unstated premise, the Separation Axiom (SA), which assumes that mining tools and the data being analyzed are epistemologically independent.
However, this premise is increasingly challenged by the emergence of LLM-driven AI agents that actively participate in data generation.
We argue that this shift leads to a new epistemic failure mode, which we term ontological self-reference, where discovered patterns may reflect the inductive biases of the tools rather than the underlying structure of the data.
This raises fundamental concerns about the credibility of knowledge discovery and the risk of self-reinforcing epistemic loops, where inferred patterns may reflect model-internal biases rather than the underlying structure of the data.
We therefore call for a rethinking of the data mining paradigm in the era of large language models, grounded in an SA-aware theoretical foundation and extended to system-level epistemic safeguards.
We argue that such a shift is necessary to conceptually rearm the discipline, preventing future knowledge discovery from collapsing into self-justifying algorithmic echo chambers.

\end{abstract}


\section{Introduction}

\subsection{From External Data to Self-Generated Data: Rethinking Independence in Data Mining Systems}




Recent advances in large language models and agent-based systems are fundamentally reshaping the data generation pipeline in modern data mining systems. 
Increasingly, data is no longer passively collected from external environments, but actively synthesized, transformed, and recorded by generative models deployed as autonomous agents.
This emerging paradigm challenges a long-standing but often implicit assumption in data mining that the process generating the data and the process analyzing the data are epistemically independent. 
We refer to this assumption as the \textbf{Separation Axiom (SA)}~\cite{Cramer1974SA}.
Under SA, mining algorithms $M$ are expected to extract patterns from data $D$ that reflect properties of the underlying generative process $G_D$, rather than the inductive biases of the model itself. We formalize this intuition as a weak independence condition between their associated belief structures: $I(\Omega_M; \Omega_{G_D}) \approx 0$, where $\Omega_M$ denotes the implicit inductive biases encoded in the mining system, and $\Omega_{G_D}$ represents the latent structure of the data-generating process.
Historically, SA has been considered a reasonable assumption due to the physical and organizational separation between data acquisition and algorithmic analysis~\cite{b1}. 
However, this separation is becoming increasingly fragile in modern AI-native systems, where the same foundation models are involved in both generating and analyzing data.
Crucially, this shift is not merely a matter of model capability, but a structural change in the data ecosystem.
The same model family may now both produce the data and extract patterns from it, leading to an emerging form of epistemic entanglement.

\subsection{Epistemic Entanglement in Agentic Data Systems}

From the deep reasoning capabilities of DeepSeek-R1 elicited via reinforcement learning~\cite{guo2025deepseek} to the fully autonomous task execution paradigm of Manus~\cite{b11}, every breakthrough in LLMs implies that more stages of the human data production pipeline are penetrated by generative AI~\cite{hagos2024recent}.
The emergence of autonomous agents~\cite{b14, b15, b16} further advances this trajectory from single-shot invocation interactions to continuous, autonomous data generation loops, where agents interact with external systems over extended periods, forming closed data lifecycles.
This leads to a systemic challenge to the Separation Axiom (SA), driven from both sides of the data ecosystem. 
On the \textbf{data side}, LLM-based agents are increasingly used in enterprises~\cite{zeng2025routine}, healthcare~\cite{wang2025survey}, and research settings~\cite{liu2025vision} to generate reports, summaries, and logs~\cite{lu2024agentlens}, which are then stored as raw data for downstream mining. 
On the \textbf{mining side}, modern data mining systems increasingly rely on LLMs as core reasoning engines for tasks such as retrieval~\cite{zhang2024ai,zhang2024agentic}, interpretation~\cite{BecharOAKH25Extracting}, and report generation~\cite{li2025unveiling}. 
As a result, both data generation and pattern extraction are increasingly mediated by the same model family.



\begin{figure}[!t]
\centering
\includegraphics[width=\linewidth]{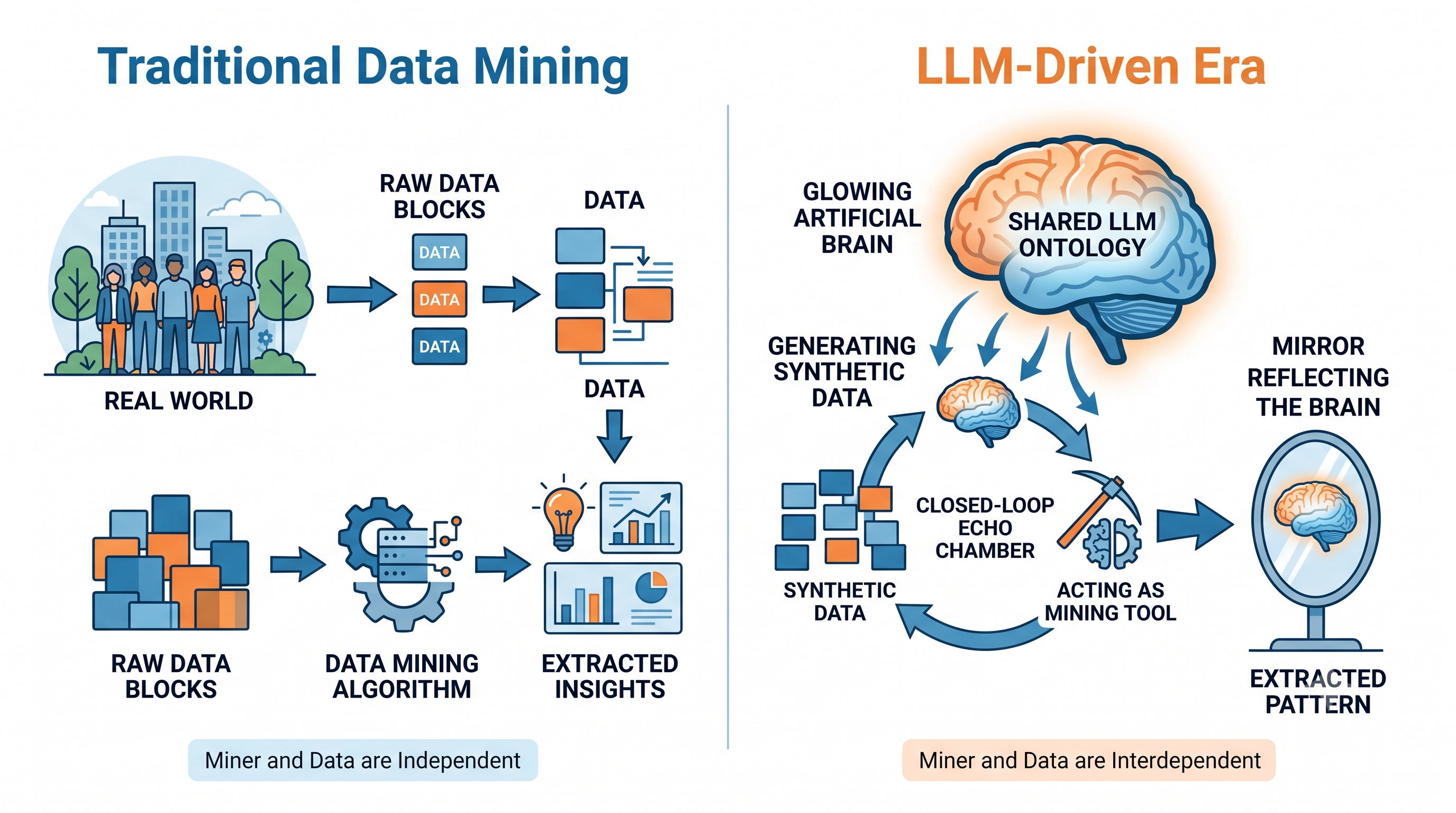}
\caption{
The collapse of the SA. Classical data mining assumes independence between data generation and mining (left). 
In AI agent-driven systems (right), both processes are coupled through a shared LLM ontology ($\Omega$), giving rise to an epistemically closed loop and ontological self-reference.
}
\label{fig:SAcollapse}
\end{figure}

When the same LLM $\theta$ is involved in both processes, i.e., $D = \text{Agent}_\theta(\text{context})$ and $P_\theta = M_\theta(D)$, the resulting patterns may reflect the internal priors of the model rather than the underlying data structure. 
This creates a form of epistemic conflation that is not explicitly addressed by classical data mining assumptions.
This epistemic entanglement undermines the reliability of classical data mining by introducing a systematic bias that propagates throughout the entire knowledge discovery pipeline.
In association rule mining, patterns extracted from agent-generated corpora reflect latent co-occurrence structures inherited from pretraining, leading to statistically plausible but misleading statistical dependencies.
A similar effect occurs in graph mining, where agent-mediated interactions can subtly distort observed topology and influence downstream inference, such as community detection.
More broadly, when agents generate system-level artifacts such as logs or compiler outputs, mining procedures recover regularities that reflect the generative biases of the underlying models rather than properties of the environment. 
These observations indicate that ontological self-reference is a general failure mode of automated data generation and analysis systems under a shared model foundation, as shown in \cref{fig:SAcollapse}.


\subsection{Our BlueSky Vision}
We envision a rethinking of data mining epistemology for the AI-native data era, where the interaction between data generation and mining systems becomes a first-class object of study rather than an implicit assumption. We outline two tightly coupled directions:

\begin{itemize}[leftmargin=*]
\item \textbf{SA-Aware Theoretical Perspective}.
Re-examining SA under modern data ecosystems where data and models are increasingly coupled, with a focus on how this coupling reshapes what we interpret as meaningful patterns.
\item \textbf{System-Level Epistemic Safeguards}.
Translating this perspective into system-level design principles for large-scale AI-native data systems to mitigate the emergence and propagation of ontological self-reference across data architectures, multi-agent ecosystems, and evaluation protocols.
\end{itemize}

\section{The Breakdown of Classical Data Mining under SA Violation}

Having established SA violation as a new failure mode of AI-native data mining, a natural question follows: can existing reliability paradigms still ensure trustworthy knowledge discovery? 
Most existing methods~\cite{b17, b18, b19} are developed under the implicit assumption that data generation and data mining maintain epistemic separation. 
Consequently, they target sources of uncertainty that are fundamentally different from those introduced by SA violation, in which the mining system and the data it analyzes become epistemically entangled~\cite{yang2025knobtunex}. 
The following discussion revisits several representative paradigms to illustrate the boundaries of their underlying assumptions in the presence of SA violation.

\subsection{Hallucination Detection: Reliability at the Output Level}
Hallucination detection~\cite{b9} is an output-level reliability paradigm that focuses on the factual correctness of individual records through retrieval augmentation, fact-checking, or other verification mechanisms. Implicitly, it assumes that improving the fidelity of each generated output leads to a more trustworthy dataset for downstream analysis.
The challenge introduced by SA violation lies at a different level of abstraction. 
Even if every generated record is factually correct, the dataset may still inherit semantic preferences and statistical regularities from the shared foundation model rather than from the underlying reality. 
Hallucination detection, therefore, verifies the correctness of individual observations, whereas SA violation concerns the epistemic fidelity of the dataset as a whole. A dataset can be entirely factual while still producing systematically misleading patterns for knowledge discovery.



\subsection{Automated Heuristic Design (AHD) Systems: Reliability through Epistemic Separation}

Recent AHD systems, including FunSearch~\cite{b2}, EoH~\cite{b3}, ReEvo~\cite{b4}, and OPRO~\cite{b5}, achieve remarkable performance by preserving epistemic separation between candidate generation and evaluation. 
LLMs propose heuristics, while independent external evaluators, such as executable sandboxes or objective optimization benchmarks, determine correctness~\cite{WangX0MXZFA24Voyager,yang2024swe}. 
Their reliability thus depends not only on model capability but also on preserving the SA.

General data mining rarely admits such an objective evaluator. 
No external sandbox can verify whether patterns mined from LLM-generated medical records, scientific literature, or social interactions faithfully reflect reality. 
The success of AHD systems therefore highlights, rather than resolves, the challenge posed by SA violation. Their reliability fundamentally depends on preserving epistemic separation through an external evaluation process, a condition that is rarely available in general data mining.
Similar assumption mismatches also exist in other paradigms, including concept drift detection~\cite{b17}, performative prediction~\cite{b7}, model collapse~\cite{b6}, and AI content watermarking~\cite{ZhaoA0W24Provable}. 
Although addressing key reliability challenges, these methods assume conditions distinct from those under SA violation.

Other work has documented declining data quality in the generative AI era. 
For instance, data feedback loops can amplify training-set biases~\cite{taori2023data}, while naive inference on synthetic tabular data can produce statistical errors and false positives~\cite{vanbreugel2023synthetic,decruyenaere2024real}.
LLMs also exhibit strong self-evaluation bias when refining their own outputs~\cite{xu2024pride}. These findings expose risks from model-generated content but remain empirical or output-level observations. 
In contrast, SA violation and the resulting ontological self-reference offer an epistemic account that unifies their underlying causes.



\section{Fundamental Challenges Beyond the Separation Axiom}

SA violation is not merely another source of error in AI-native data mining. 
By dissolving the separation between data generation and mining, it undermines the assumptions governing how knowledge is formed, propagated, and validated throughout the discovery lifecycle.



\subsection{Losing External Ground Truth}

Explainability assumes that explanations can ultimately be grounded in observations independent of the model itself~\cite{ZhangZHQW25Customized}.
Under SA violation, however, this assumption no longer holds. 
In AI-native data systems, the entire epistemic chain can be written as
\begin{equation}
    D = \mathrm{Agent}_{\theta}(\mathrm{context}) \rightarrow P = M_{\theta}(D) \rightarrow E = \mathrm{LLM}_{\theta}(P,D),
\end{equation}
where data generation, pattern discovery, and explanation all originate from the same underlying family of models.
As a result, explanations can remain internally consistent without an external reference for validating whether the discovered patterns reflect reality or merely the model's own priors. 
Explainability therefore becomes a property of the entire knowledge discovery pipeline rather than of the model alone.
This, in turn, calls for a new theory of explainability under epistemic coupling.



\subsection{Collective Epistemic Reinforcement}

Modern AI-native systems increasingly comprise multiple interacting agents that continuously exchange information through shared context, retrieved memories, and generated outputs~\cite{zhang2025survey}. 
Under SA violation, each agent repeatedly consumes the outputs of others as context and generates new observations, forming a closed feedback loop within the shared data ecosystem. 
Because these agents share the same epistemic foundation, beliefs inherited from the underlying models are repeatedly propagated and reinforced. 
As a result, initially biased beliefs can gradually accumulate statistical support and eventually emerge as reliable consensus despite lacking external grounding.
Unlike model collapse, this reinforcement occurs entirely through context exchange rather than model updates. 
Under SA violation, consensus alone can no longer be treated as evidence of reliability.


\subsection{Losing Independent Evaluation}
Traditional evaluation methodologies rely on the assumption that agreement between model outputs and observed data reflects external validity. 
Under SA violation, however, both data and learned patterns are shaped by the same epistemic foundation. 
Evaluation therefore no longer compares predictions against independent evidence, but instead measures internal consistency within a shared epistemic system.
Consequently, statistical significance, reproducibility, and generalization cease to serve as reliable indicators of external validity, becoming measures of epistemic self-consistency instead.


These three challenges indicate that SA violation induces a unified epistemic collapse in data mining systems. 
The core issue is not failure at individual stages, but the disappearance of epistemic independence across the entire pipeline of data generation, pattern discovery, and evaluation. 
This collapse motivates the need for a fundamentally new theory for reasoning about credibility under epistemically coupled systems.

\section{The BlueSky Initiative: Toward Epistemically Aware Data Mining Systems}

Preventing epistemic collapse requires coordinated mechanisms across data, models, and evaluation, rather than isolated interventions at individual stages.

\subsection{Stage I: Epistemic Observability}
A first requirement is to make epistemic coupling explicitly measurable rather than implicitly assumed. 
We introduce an epistemic separation signal, termed the Excess Alignment Ratio $S(M,D \mid D_0)$. 
This signal quantifies the degree of alignment between model-induced priors and data-generating structures relative to a baseline. 
Unlike conventional distributional metrics, $S(M,D \mid D_0)$ captures whether the observed dataset is shaped by the same epistemic foundation as the mining model itself.
Operationally, this notion can be instantiated via belief fingerprint probing. 
A standardized probe set $Q = \{q_1, ..., q_k\}$ is constructed to span critical propositions in the target domain. 
Both the model $M$ and dataset $D$ are mapped into comparable distributions over this probe space, yielding a belief fingerprint $F_M$ and a generative fingerprint $F_D$.
The epistemic separation is then defined as
\begin{equation}\label{eq:s}
S(M, D \mid D_0) = \min\left(1, \frac{\bar{W}(F_M, F_D)}{\bar{W}(F_M, F_{D_0}) + \epsilon}\right),
\end{equation}
where $\bar{W}(\cdot)$ denotes a normalized distributional distance such as Wasserstein, $D_0$ is a set of baseline objective observations, and $\epsilon$ is a small constant for numerical stability.
A high value of $S$ indicates weak epistemic coupling between model and data, whereas a low value indicates strong alignment induced by shared epistemic priors.
This formulation makes epistemic entanglement explicitly measurable, providing a foundation for downstream intervention and stabilization mechanisms.

\subsection{Stage II: Epistemic Intervention}

Once epistemic coupling becomes observable through $S(M,D)$, the challenge of data mining shifts from estimating patterns under a fixed dataset to reasoning under a variable epistemic structure induced by shared model priors.
A central implication of this shift is that validity can no longer be defined purely in terms of fit to the observed data. Instead, it must account for how both data and patterns co-vary under changes in the underlying epistemic assumptions $\Omega_M$. 
This leads to a stability-based notion of inference, where a pattern is considered reliable if it remains invariant under epistemic perturbations~\cite{arjovsky2019invariant,ahuja2020invariant,WangLZ25OOD}.
Formally, this can be understood through controlled variations of the model-induced epistemic structure, such as modifying prompts or model configurations, and observing the induced changes in both data representation and extracted pattern $P$. 
Patterns that exhibit high stability under such perturbations are more likely to reflect structure in the external environment rather than artifacts of shared prior.
In parallel, epistemic bias in the observed dataset can be partially addressed by adjusting the influence of individual samples according to their consistency with the model's generative tendency $P_M(x)$. 
Rather than treating this as a correction toward an unknown ground truth, it serves as a rebalancing mechanism that reduces over-amplification of model-consistent observations while preserving epistemically diverse signals.
This establishes a unified intervention principle: under SA violation, the objective of data mining is no longer to eliminate bias, but to identify patterns that are robust to epistemic variation.

\subsection{Stage III: Epistemic Stabilization}

\begin{figure}[!t]
\centering
\includegraphics[width=\linewidth]{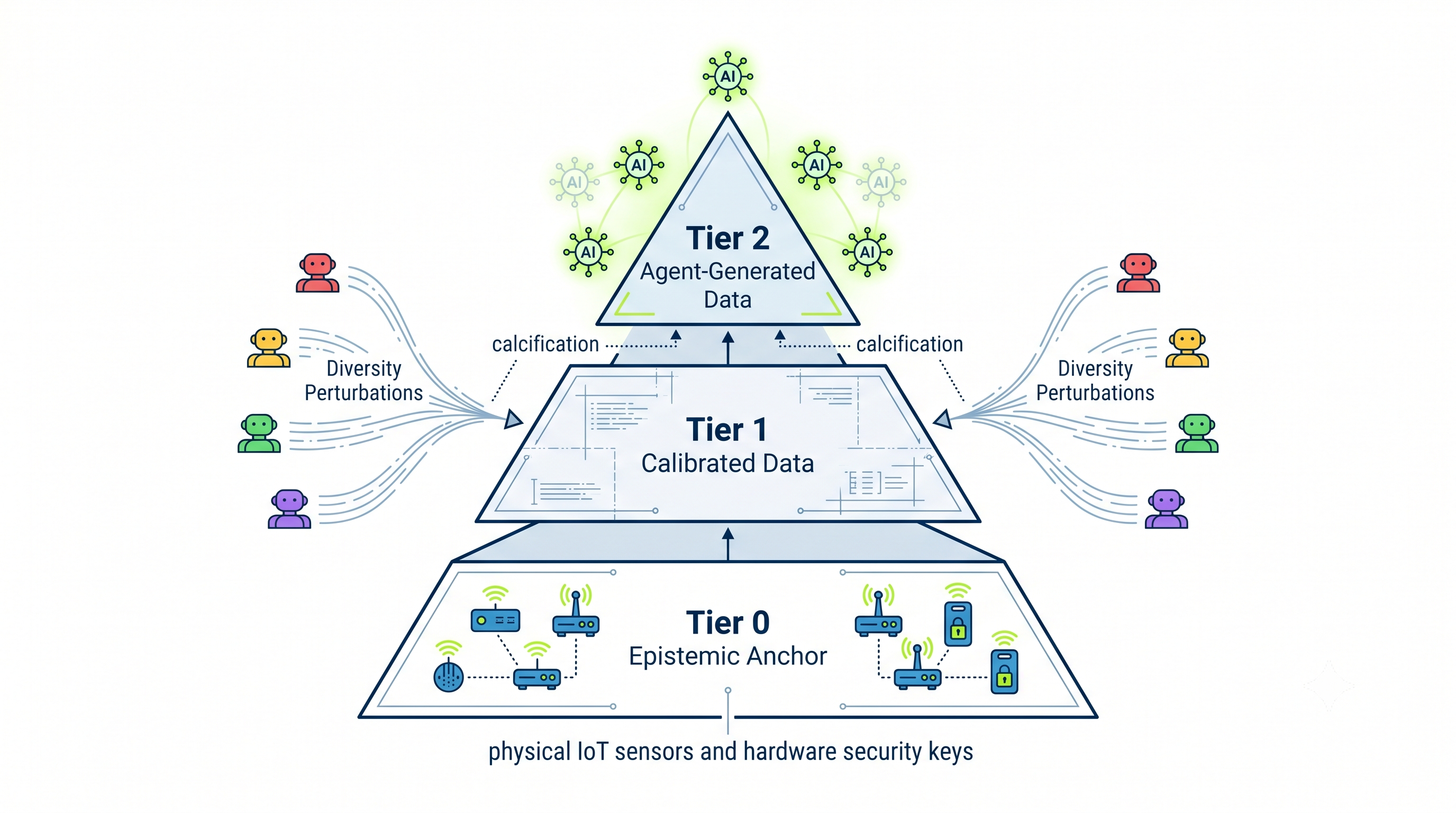}
\caption{The Tri-Tier Epistemic Anchoring (TEA) Architecture. A system-level framework for representing and managing epistemic stratification in AI-native data systems.
}
\label{fig:TEA}
\end{figure}

The theoretical constructs introduced above require system-level instantiation to operate under real-world AI-native data pipelines. 
Unlike conventional system design, the objective is not modular enhancement, but the prevention of epistemic closure across data generation, interaction, and evaluation.
To this end, we introduce a Tri-Tier Epistemic Anchoring (TEA) architecture that makes the latent epistemic structure of data explicitly representable. 
Rather than treating datasets as homogeneous inputs, TEA partitions data according to their degree of entanglement with model-induced priors. 
Tier 0 corresponds to raw observational anchors that approximate external reality; Tier 1 contains data partially transformed by models while retaining traceability to Tier 0; Tier 2 corresponds to fully model-generated content. 
In practice, even nominally human-generated data may belong to higher tiers due to pervasive model-assisted production, making epistemic contamination a structural property rather than an exception.
The architecture is illustrated in \cref{fig:TEA}.

This stratified representation of data directly affects how information propagates through multi-agent systems. 
When agents share a common foundation model, their outputs are repeatedly reintroduced into subsequent contexts through iterative multi-agent interaction~\cite{wu2023autogen,hong2024metagpt}, forming a reinforcement loop over shared epistemic priors. 
Such feedback dynamics tend to concentrate the system toward a single epistemic attractor. 
Epistemic diversity injection mitigates this effect by incorporating agents built on heterogeneous model families or training corpora. 
These agents act as controlled perturbations that interrupt convergence dynamics. 
Disagreement among heterogeneous agents when evaluating identical patterns thus becomes a diagnostic signal of potential epistemic artifacts, revealing instability otherwise masked by consensus formation.

Evaluation must then be interpreted under this same epistemic structure. 
The validity of standard evaluation metrics such as accuracy, F1, and reproducibility relies on epistemic independence between data and model.
Once this assumption fails, evaluation no longer measures external validity, but only internal consistency within a shared epistemic system. 
A more appropriate criterion is stability under epistemic perturbation: whether a discovered pattern remains invariant across changes in data provenance, model family, or inference context. 
Evaluation therefore shifts from fixed-dataset agreement toward invariance under epistemic reconfiguration.
Epistemic validity in AI-native data mining is determined by a coupled control regime in which observability, invariance, and stratification jointly characterize the conditions under which inference remains externally grounded, rather than collapsing into self-referential consistency.

\subsection{Main Claim}

Under SA violation, data mining is characterized by epistemic non-identifiability between data generation and inference, in which external structure and model-induced structure become observationally indistinguishable.
The BlueSky framework does not attempt to restore classical separation. Instead, it formalizes epistemic entanglement through observable separation signals, replaces validity with invariance under epistemic perturbation, and enforces system-level stratification to prevent collapse into self-referential consistency.

\section{Conclusion}


Long implicit in modern data mining, the Separation Axiom can no longer be taken for granted as AI-native data systems proliferate. We argue that epistemic coupling between data generation and mining creates a new failure mode not adequately captured by existing accounts of distribution shift, explainability, or statistical evaluation.
Rather than viewing this challenge as another problem to be mitigated, we suggest that it calls for a broader rethinking of the foundations of data mining. The BlueSky initiative outlined in this paper provides one possible direction by making epistemic coupling observable, enabling stability-oriented intervention, and promoting system-level epistemic safeguards.
More broadly, we hope this work encourages the community to treat epistemic separation as an explicit design principle for future AI-native data mining systems. As data is increasingly generated, curated, and interpreted by intelligent agents, understanding \emph{how knowledge is produced} may become as important as improving \emph{what patterns can be discovered}.



\clearpage
\bibliographystyle{IEEEtran}
\bibliography{reference}

@article{b1,
  title={From data mining to knowledge discovery in databases},
  author={Fayyad, Usama M and Piatetsky-Shapiro, Gregory and Smyth, Padhraic},
  journal={AI magazine},
  volume={17},
  number={3},
  pages={37--54},
  year={1996}
}

@article{b2,
  title={Mathematical discoveries from program search with large language models},
  author={Romera-Paredes, Bernardino and others},
  journal={Nature},
  volume={625},
  pages={468--475},
  year={2024}
}

@inproceedings{b3,
  title={Evolution of heuristics: Towards efficient automatic algorithm design using large language model},
  author={Liu, F and others},
  booktitle={Proceedings of the International Conference on Machine Learning (ICML)},
  year={2024}
}

@inproceedings{b4,
  title={ReEvo: Large language models as hyper-heuristics with reflective evolution},
  author={Ye, H and others},
  booktitle={Proceedings of the Conference on Neural Information Processing Systems (NeurIPS)},
  year={2024}
}

@inproceedings{b5,
  title={Large language models as optimizers},
  author={Yang, C and others},
  booktitle={Proceedings of the International Conference on Learning Representations (ICLR)},
  year={2024}
}

@article{b6,
  title={{AI} models collapse when trained on recursively generated data},
  author={Shumailov, I and others},
  journal={Nature},
  volume={631},
  pages={755--759},
  year={2024}
}

@inproceedings{b7,
  title={Performative prediction},
  author={Perdomo, J C and Zrnic, T and Mendler-Dunner, C and Hardt, M},
  booktitle={Proceedings of the International Conference on Machine Learning (ICML)},
  year={2020}
}

@article{b9,
  title={Survey of hallucination in natural language generation},
  author={Ji, Z and others},
  journal={ACM Computing Surveys},
  volume={55},
  number={12},
  pages={248},
  year={2023}
}

@article{b11,
  title={From Mind to Machine: The Rise of Manus AI as a Fully Autonomous Digital Agent},
  author={Shen, M and Li, Y and Chen, L and Fan, Z and Li, Y and Yang, Q},
  journal={arXiv preprint arXiv:2505.02024},
  year={2025}
}

@inproceedings{b14,
  title={Generative agents: Interactive simulacra of human behavior},
  author={Park, J S and others},
  booktitle={Proceedings of the Symposium on User Interface Software and Technology (UIST)},
  year={2023}
}

@article{b15,
  title={A survey on large language model based autonomous agents},
  author={Wang, X and others},
  journal={Frontiers of Computer Science},
  volume={18},
  number={6},
  pages={186345},
  year={2024}
}

@inproceedings{b16,
  title={AgentBench: Evaluating LLMs as agents},
  author={Wei, P and others},
  booktitle={Proceedings of the International Conference on Learning Representations (ICLR)},
  year={2024}
}

@inproceedings{b17,
  title={AnDri: Adaptive anomaly detection in the presence of concept drift},
  author={Park, Jongjun and Chiang, Fei and Milani, Mostafa},
  booktitle={Proceedings of the ACM International Conference on Information and Knowledge Management (CIKM)},
  year={2025}
}

@article{b18,
  title={CleanPatrick: A benchmark for image data cleaning},
  author={Gr{\"o}ger, Fabian and Lionetti, Simone and Gottfrois, Philippe and Gonzalez-Jimenez, Alvaro and others},
  journal={arXiv preprint arXiv:2505.11034},
  year={2025}
}

@inproceedings{b19,
  title={From Data Imputation to Data Cleaning-Automated Cleaning of Tabular Data Improves Downstream Predictive Performance},
  author={J{\"a}ger, Sebastian and Biessmann, Felix},
  booktitle={Proceedings of the International Conference on Artificial Intelligence and Statistics (AISTATS)},
  year={2024}
}

@inproceedings{ZhaoA0W24Provable,
  author       = {Xuandong Zhao and Prabhanjan Vijendra Ananth and
                  Lei Li and Yu{-}Xiang Wang},
  title        = {Provable Robust Watermarking for AI-Generated Text},
  booktitle    = {The International Conference on Learning Representations ({ICLR})},
  year         = {2024}
}

@inproceedings{li2025unveiling,
  title={Unveiling the merits and defects of {LLMs} in automatic review generation for scientific papers},
  author={Li, Ruochi and Zhang, Haoxuan and Gehringer, Edward and Xiao, Ting and Ding, Junhua and Chen, Haihua},
  booktitle={IEEE International Conference on Data Mining (ICDM)},
  pages={1370--1379},
  year={2025},
}

@inproceedings{yang2025knobtunex,
  title={KnobTuneX: LLM-Enhanced Automatic Database Tuning via Structured Reasoning},
  author={Yang, Jianwen and Zhang, Qiuhong and Xu, Xinrun and Wu, Yurong and Zhang, Shuo and Ding, Zhiming},
  booktitle={IEEE International Conference on Data Mining (ICDM)},
  pages={843--852},
  year={2025},
  organization={IEEE}
}

@inproceedings{ZhangZHQW25Customized,
  author       = {Haichao Zhang and Chong Zhang and Peiyu Hu and Shi Qiu and Jia Wang},
  title        = {Customized Retrieval-Augmented Generation with {LLM} for Debiasing Recommendation Unlearning},
  booktitle    = {{IEEE} International Conference on Data Mining, ({ICDM})},
  pages        = {1702--1711},
  year         = {2025},
}

@article{Cramer1974SA, 
title={A Definition of Separation Axiom}, 
volume={17}, 
DOI={10.4153/CMB-1974-086-8}, 
number={4}, 
journal={Canadian Mathematical Bulletin}, 
author={Cramer, T.}, 
year={1974}, 
pages={485–491}
}

@article{zeng2025routine,
  title={Routine: A structural planning framework for llm agent system in enterprise},
  author={Zeng, Guancheng and Chen, Xueyi and Hu, Jiawang and Qi, Shaohua and Mao, Yaxuan and Wang, Zhantao and Nie, Yifan and Li, Shuang and Feng, Qiuyang and Qiu, Pengxu and others},
  journal={arXiv preprint arXiv:2507.14447},
  year={2025}
}

@article{wang2025survey,
  title={A survey of llm-based agents in medicine: How far are we from baymax?},
  author={Wang, Wenxuan and Ma, Zizhan and Wang, Zheng and Wu, Chenghan and Ji, Jiaming and Chen, Wenting and Li, Xiang and Yuan, Yixuan},
  journal={Findings of the Association for Computational Linguistics (ACL)},
  pages={10345--10359},
  year={2025}
}

@article{liu2025vision,
  title={A vision for auto research with llm agents},
  author={Liu, Chengwei and Wang, Chong and Cao, Jiayue and Ge, Jingquan and Wang, Kun and Zhang, Lyuye and Cheng, Ming-Ming and Zhao, Penghai and Li, Tianlin and Jia, Xiaojun and others},
  journal={arXiv preprint arXiv:2504.18765},
  year={2025}
}

@article{lu2024agentlens,
  title={Agentlens: Visual analysis for agent behaviors in llm-based autonomous systems},
  author={Lu, Jiaying and Pan, Bo and Chen, Jieyi and Feng, Yingchaojie and Hu, Jingyuan and Peng, Yuchen and Chen, Wei},
  journal={IEEE Transactions on Visualization and Computer Graphics (TVCG)},
  year={2024},
  publisher={IEEE}
}

@inproceedings{zhang2024ai,
  title={Ai agent for information retrieval: Generating and ranking},
  author={Zhang, Yongfeng and Liu, Zhiwei and Wen, Qingsong and Pang, Linsey and Liu, Wei and Yu, Philip S},
  booktitle={Proceedings of the 33rd ACM International Conference on Information and Knowledge Management},
  pages={5605--5607},
  year={2024}
}

@article{zhang2024agentic,
  title={Agentic information retrieval},
  author={Zhang, Weinan and Liao, Junwei and Li, Ning and Du, Kounianhua and Lin, Jianghao},
  journal={arXiv preprint arXiv:2410.09713},
  year={2024}
}

@inproceedings{BecharOAKH25Extracting,
  author       = {Amine Bechar and Adel Oulefki and Abbes Amira and
                  Fatih Kurogollu and Yassine Himeur},
  title        = {Extracting Actionable Insights from Building Energy Data using Vision LLMs on Wavelet and 3D Recurrence Representations},
  booktitle    = {{IEEE} International Conference on Data Mining ({ICDM})},
  pages        = {1040--1047},
  publisher    = {{IEEE}},
  year         = {2025},
}

@article{guo2025deepseek,
  title={DeepSeek-R1 incentivizes reasoning in LLMs through reinforcement learning},
  author={Guo, Daya and Yang, Dejian and Zhang, Haowei and Song, Junxiao and Wang, Peiyi and Zhu, Qihao and Xu, Runxin and Zhang, Ruoyu and Ma, Shirong and Bi, Xiao and others},
  journal={Nature},
  volume={645},
  number={8081},
  pages={633--638},
  year={2025},
  publisher={Nature Publishing Group UK London}
}

@article{hagos2024recent,
  title={Recent advances in generative {AI} and large language models: Current status, challenges, and perspectives},
  author={Hagos, Desta Haileselassie and Battle, Rick and Rawat, Danda B},
  journal={IEEE Transactions on Artificial Intelligence (TAI)},
  volume={5},
  number={12},
  pages={5873--5893},
  year={2024},
}

@inproceedings{WangLZ25OOD,
  author       = {Yuanchao Wang and Zhao{-}Rong Lai and
                  Tianqi Zhong},
  title        = {Out-of-distribution Generalization for Total Variation based Invariant Risk Minimization},
  booktitle    = {The Thirteenth International Conference on Learning Representations ({ICLR})},
  year         = {2025}
}

@article{arjovsky2019invariant,
  title={Invariant risk minimization},
  author={Arjovsky, Martin and Bottou, L{\'e}on and Gulrajani, Ishaan and Lopez-Paz, David},
  journal={arXiv preprint arXiv:1907.02893},
  year={2019}
}

@inproceedings{ahuja2020invariant,
  title={Invariant risk minimization games},
  author={Ahuja, Kartik and Shanmugam, Karthikeyan and Varshney, Kush and Dhurandhar, Amit},
  booktitle={International Conference on Machine Learning ({ICML})},
  pages={145--155},
  year={2020},
}

@article{wu2023autogen,
  title={Autogen: Enabling next-gen llm applications via multi-agent conversation},
  author={Wu, Qingyun and Bansal, Gagan and Zhang, Jieyu and Wu, Yiran and Li, Beibin and Zhu, Erkang and Jiang, Li and Zhang, Xiaoyun and Zhang, Shaokun and Liu, Jiale and others},
  journal={arXiv preprint arXiv:2308.08155},
  year={2023}
}

@inproceedings{hong2024metagpt,
  title={MetaGPT: Meta programming for a multi-agent collaborative framework},
  author={Hong, Sirui and Zhuge, Mingchen and Chen, Jonathan and Zheng, Xiawu and Cheng, Yuheng and Wang, Jinlin and Zhang, Ceyao and Yau, Steven and Lin, Zijuan and Zhou, Liyang and others},
  booktitle={International Conference on Learning Representations ({ICLR})},
  volume={2024},
  pages={23247--23275},
  year={2024}
}

@article{zhang2025survey,
  title={A survey on the memory mechanism of large language model-based agents},
  author={Zhang, Zeyu and Dai, Quanyu and Bo, Xiaohe and Ma, Chen and Li, Rui and Chen, Xu and Zhu, Jieming and Dong, Zhenhua and Wen, Ji-Rong},
  journal={ACM Transactions on Information Systems},
  volume={43},
  number={6},
  pages={1--47},
  year={2025},
}

@article{WangX0MXZFA24Voyager,
  author       = {Guanzhi Wang and Yuqi Xie and Yunfan Jiang and
                  Ajay Mandlekar and Chaowei Xiao and Yuke Zhu and Linxi Fan and
                  Anima Anandkumar},
  title        = {Voyager: An Open-Ended Embodied Agent with Large Language Models},
  journal      = {Transactions on Machine Learning Research (TMLR},
  year         = {2024},
}

@article{yang2024swe,
  title={Swe-agent: Agent-computer interfaces enable automated software engineering},
  author={Yang, John and Jimenez, Carlos and Wettig, Alexander and Lieret, Kilian and Yao, Shunyu and Narasimhan, Karthik and Press, Ofir},
  journal={Advances in Neural Information Processing Systems ({NeurIPS})},
  volume={37},
  pages={50528--50652},
  year={2024}
}

@inproceedings{taori2023data,
  title={Data Feedback Loops: Model-driven Amplification of Dataset Biases},
  author={Taori, Rohan and Hashimoto, Tatsunori B},
  booktitle={Proceedings of the 40th International Conference on Machine Learning (ICML)},
  volume={202},
  pages={33883--33920},
  year={2023},
  organization={PMLR}
}

@inproceedings{vanbreugel2023synthetic,
  title={Synthetic Data, Real Errors: How (Not) to Publish and Use Synthetic Data},
  author={van Breugel, Boris and Qian, Zhaozhi and van der Schaar, Mihaela},
  booktitle={Proceedings of the 40th International Conference on Machine Learning (ICML)},
  volume={202},
  pages={34793--34808},
  year={2023},
  organization={PMLR}
}

@inproceedings{decruyenaere2024real,
  title={The Real Deal Behind the Artificial Appeal: Inferential Utility of Tabular Synthetic Data},
  author={Decruyenaere, Alexander and Dehaene, Heidelinde and Rabaey, P and Polet, C and Decruyenaere, J and Vansteelandt, S and Demeester, T},
  booktitle={Proceedings of the 40th Conference on Uncertainty in Artificial Intelligence (UAI)},
  volume={244},
  pages={966--996},
  year={2024},
  organization={PMLR}
}

@inproceedings{xu2024pride,
  title={Pride and Prejudice: LLM Amplifies Self-Bias in Self-Refinement},
  author={Xu, Wenda and Zhu, Guanglei and Zhao, Xuandong and Pan, Liangming and Li, Lei and Wang, William Yang},
  booktitle={Proceedings of the 62nd Annual Meeting of the Association for Computational Linguistics (Volume 1: Long Papers)},
  pages={15474--15492},
  year={2024}
}

\end{document}